\documentclass[aps,prl,preprint,superscriptaddress]{revtex4}
\usepackage{graphicx}
\usepackage{times}
\usepackage{amsmath}
\usepackage{textcomp}
\usepackage{bm}
\usepackage{braket}

\begin{document}

\title{Magnetoresistance from broken spin helicity}
\author{D.P. Leusink}
\affiliation{Faculty of Science and Technology and MESA+ Institute for Nanotechnology, University of Twente, The Netherlands}
\author{R.G.J. Smits}
\affiliation{Faculty of Science and Technology and MESA+ Institute for Nanotechnology, University of Twente, The Netherlands}
\author{P. Ngabonziza}
\affiliation{Faculty of Science and Technology and MESA+ Institute for Nanotechnology, University of Twente, The Netherlands}
\author{X.L. Wang}
\affiliation{Institute for Superconducting and Electronic Materials, University of Wollongong, Wollongong, Australia }
\author{S. Wiedmann}
\affiliation{High Field Magnet Laboratory, Institute of Molecules and Materials, Radboud University Nijmegen, The Netherlands}
\author{U. Zeitler}
\affiliation{High Field Magnet Laboratory, Institute of Molecules and Materials, Radboud University Nijmegen, The Netherlands}
\author{A. Brinkman}
\affiliation{Faculty of Science and Technology and MESA+ Institute for Nanotechnology, University of Twente, The Netherlands}
\date{\today}

\begin{abstract}
\textbf{The propensity of some materials and multilayers to have a magnetic field dependent resistance, called magnetoresistance, has found commercial applications such as giant magnetoresistance harddisk read heads. But magnetoresistance can also be a powerful probe of electronic and magnetic interactions in matter. For example, magnetoresistance can be used to analyze multiband conductivity, conduction inhomogeneity, localized magnetic moments, and (fractional) Landau level structure. For materials with strong spin-orbit interaction, magnetoresistance can be used as a probe for weak antilocalization or a nontrivial Berry phase, such as in topological insulator surface states. For the three dimensional topological insulators a large and linear magnetoresistance is often used as indication for underlying non-trivial topology, although the origin of this effect has not yet been established.\\
Here, we observe a large magnetoresistance in the conducting bulk state of Bi$_2$Te$_3$. We show that this type of large magnetoresistance is due to the competition between helical spin-momentum locking (i.e. spin rotates with momentum direction) and the unidirectional spin alignment by an applied magnetic field. Warping effects are found to provide the (quasi) linear dependence on magnetic field. We provide a quantitative model for the helicity breaking induced magnetoresistance that can be applied to a vast range of materials, surfaces or interfaces with weak to strong spin-orbit interactions, such as the contemporary oxide interfaces, bulk Rashba systems, and topological insulator surface states.}
\end{abstract}
\maketitle

Linear magnetoresistance studies go back as far as almost a century, such as Kapitza's magnetoresistance observations in Bi \cite{Kapitza} and in metals with large open Fermi surfaces. Inhomogeneities in disordered conductors (such as regions with higher conductivity than others) can create deviations of the direction of the applied current. Classical resistor network models \cite{Meera2003,Meera2005} show that in this case a Hall voltage admixture into the longitudinal voltage occurs, giving rise to the so-called classical linear magnetoresistance. Abrikosov derived in 1969 a linear magnetoresistance when only one Landau level is filled \cite{Abrikosov1969,Abrikosov2000}. The conditions for the sole occupation of the lowest Landau level is that both the Fermi energy, $E_F$, and $k_BT$ are much smaller than the energy difference between the lowest and next Landau level. This is called the extreme quantum limit and the resulting linear magnetoresistance is called quantum linear magnetoresistance. Typically, the conditions for the extreme quantum limit are only fulfilled for narrow gap semiconductors or semi-metals with very small Fermi pockets and very low effective masses. Examples could be Bi \cite{Kapitza} and n-type doped InSb \cite{Hu}. A quantum linear magnetoresistance was also suggested for $\beta$-Ag$_2$Te \cite{Xu} since the calculated linear dispersion relation would typically relax the conditions for the extreme quantum limit \cite{Zhang}.

The large and sometimes linear magnetoresistance in Bi-based and Heusler topological insulators \cite{He,Qu,Tang,Wang,WWang} has also been linked to the fact that topological insulators have a linear dispersion relation. However, few experiments fulfill the extreme quantum limit criterion. Additionally, it was suggested \cite{Moodera} that the logarithmic magnetic field dependence of weak antilocalization in topological insulators might combine with a quadratic magnetic field dependence, rendering an effective total dependence that is quasi-linear. But this mechanism can only explain a correction to conductance on the order of the conductance quantum $\frac{2e^2}{h}=(13.9$ k$\Omega)^{-1}$.

Here, we propose an alternative mechanism that is applicable to a wide class of systems with spin-orbit coupling (from weak to strong), either at surfaces, interfaces or in the bulk. The concept is most easily illustrated with a topological insulator. The spin of the electrons in the Dirac cone of a topological surface state is locked to momentum in a helical way, as measured by angle resolved photoemission spectroscopy \cite{Hsieh}, and as illustrated in Fig. 1a. The electron spin has opposite helicity above and below the Dirac point. When scattering is considered to a state with opposite momentum, the spin state would be exactly orthogonal. The projection of initial and final states are then zero, which is the underlying physics for the topological protection against backscattering in these systems and for the presence of weak antilocalization \cite{WAL} with the associated Berry phase of $\pi$. Finite angle scattering other than $\theta=\pi$ is allowed, albeit with a suppression factor of $\frac{1}{2}\left(1+\cos \theta \right)$ (see below for details). In a Boltzmann scattering picture, forward scattering does not contribute to resistance, which gives another factor of $1-\textrm{cos }\theta$. Integrating the product of the scatter and Boltzmann factors over all angles gives a total factor of 4 reduction in the resistance when compared to the situation without topological protection \cite{Armitage}. An external field perpendicular to a topological surface state opens a gap at the Dirac point and tends to align the spins along the applied field direction due to the Zeeman interaction, see Fig. 1b. The helical spin symmetry gets broken. The spin states thereby become less orthogonal and scattering is enhanced until the field is so large that the underlying helical spin symmetry is completely absent and the same factor of 4 increase in resistance appears.
\begin{figure}
\includegraphics[width=0.6\textwidth]{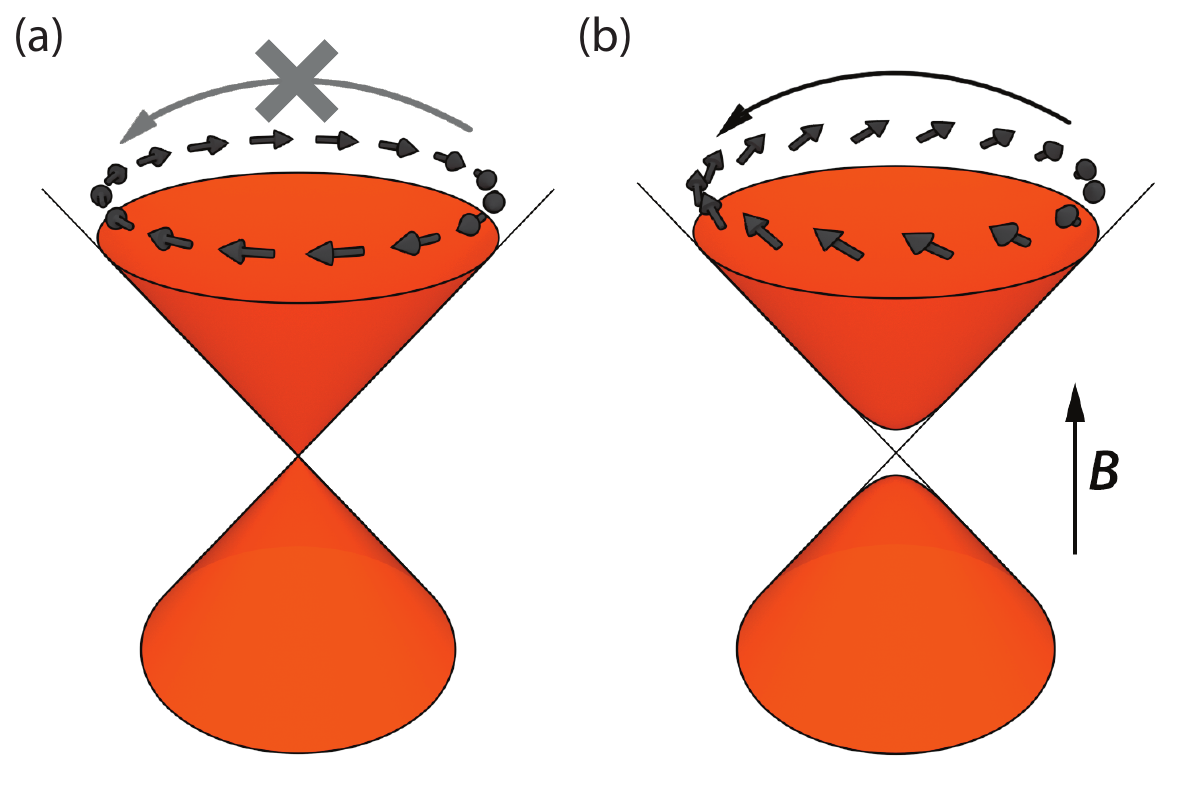}
\caption{\textbf{| Enhanced scattering from broken spin helicity.} (a) A topological insulator Dirac cone at zero field with full spin helical symmetry. Backscattering is not possible because of the orthogonal spin states. (b) The topological Dirac cone in a perpendicular magnetic field has a band gap at the Dirac point and a broken helical symmetry due to the out-of-plane alignment of the spins. Backscattering is now possible because of overlap between the spin states.}  
\label{fig:1}
\end{figure}

As shown below, the concept of helical magnetoresistance in a topological insulator can be generalized to two-dimensional electron gases with quadratic dispersion and even to three-dimensional conduction. In fact, we will show that the helical magnetoresistance effect is present in the bulk conduction of a Hall bar made of Bi$_2$Te$_3$. The helical magnetoresistance can be quantitatively fitted by a straightforward Boltzmann model. We find that Fermi surface warping effectively acts as a magnetic field offset, giving rise to a quasi-linear dependence of the resistance on the magnetic field. We derive an analytical expression that can be used to apply this magnetoresistance concept easily to many other systems.

The Bi$_2$Te$_3$ Hall device of Fig. 2 is structured using e-beam lithography. We deposited flakes of Bi$_2$Te$_3$ crystal onto silicon substrates by mechanical exfoliation. The gold contacts are sputter deposited and the flake is structured into a Hall bar using Ar ion beam etching. A scanning electron microscopy image of the device is shown in Fig. 2a and a cross section of the device in Fig. 2b. 
The longitudinal resistance at low temperatures shows Shubnikov-de Haas oscillations (see the Supplementary Information), from which we obtain a surface carrier density of $1.2 \times 10^{12} \textrm{ cm}^{-2}$. The Hall coefficient, $1/eR_H= 5.6 \times 10^{14}$ cm$^{-2}$ indicates a much larger carrier density, which should be interpreted as bulk carriers, $n_{3D}=n_{2D}/d = 2.7 \times 10^{19}$ cm$^{-3}$, where the flake thickness $d =$ 210 nm is determined by atomic force microscopy. The surface to bulk ratio of the carrier densities is consistent with similar devices from the same crystal \cite{Menno}. At low temperatures the surface conduction is enhanced because of the higher mobility but for temperatures above 45 K we estimate the surface conduction contribution to be negligible (see the Supplementary Information). The measured magnetoresistance, shown for 45 K in Fig. 2c for different magnetic field directions, can therefore be safely attributed to the bulk conduction channel. Figure 2d shows the longitudinal and transverse conductivity as deduced from the longitudinal resistivity and transverse Hall signal.

\begin{figure}
\includegraphics[width=\textwidth]{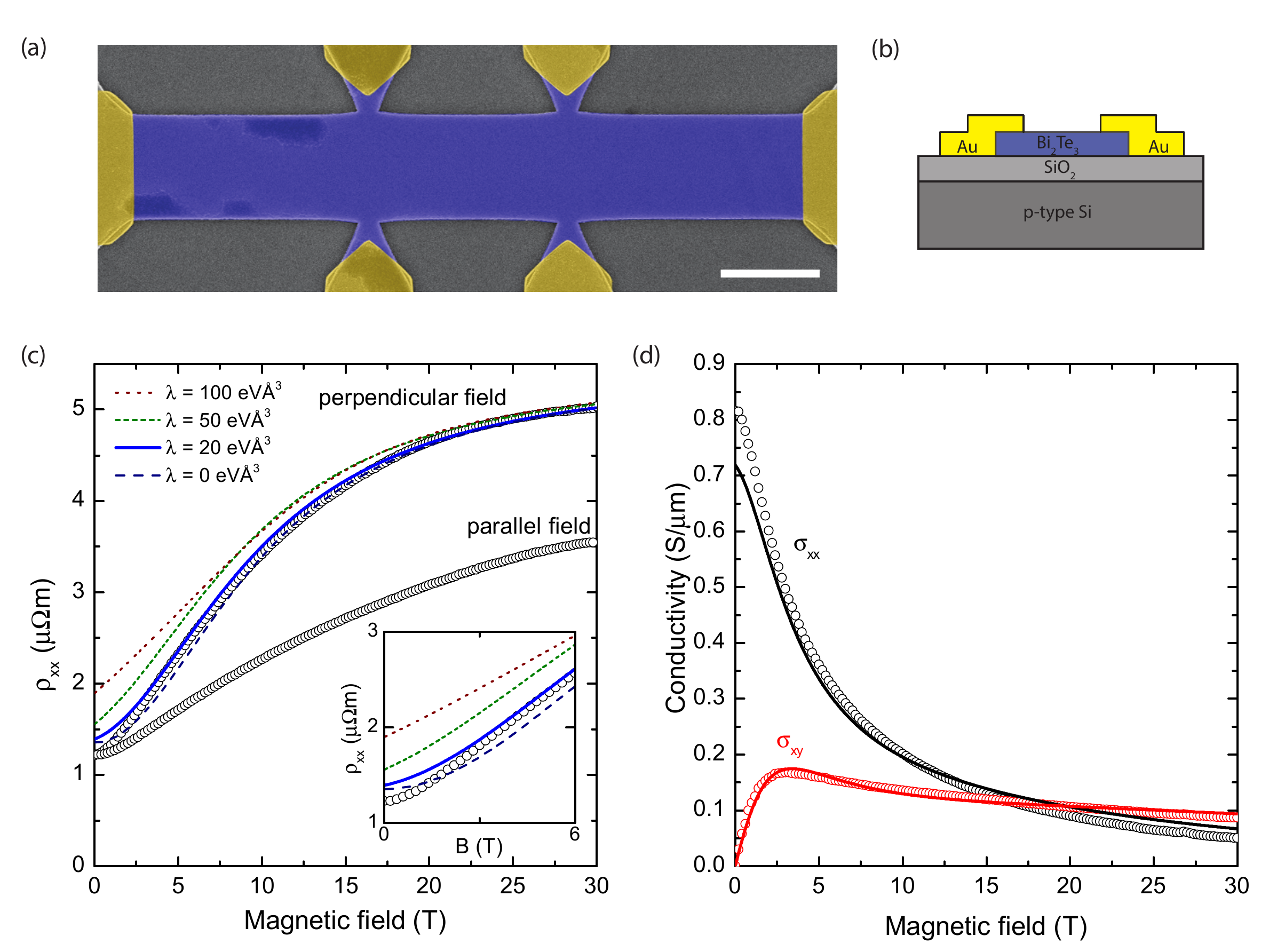}
\caption{\textbf{| Helical magnetoresistance device.} (a) Scanning electron microscopy image (false color) of the Bi$_2$Te$_3$ Hall bar. The scale bar indicates 5 \textmu m. (b) Schematic cross-section of the device. (c) Magnetic field dependent longitudinal resistivity for different orientations of the magnetic field. The inset shows the effect of adding a warping term to the fitting, rendering the magnetoresistance linear at low fields. (d) Longitudinal and transverse conductivity data obtained from the measured longitudinal and transverse resistance, simultaneously fitted with the helical magnetoresistance model.}  
\label{fig:2}
\end{figure}

The measured magnetoresistance starts off with a small quadratic dependence on magnetic field but is then characterized by a quasi-linear increase to a magnetic field of about 15 T after which saturation to a constant resistance appears. Kapitza's linear magnetoresistance model \cite{Kapitza} does not apply here, since open Fermi surfaces require large Fermi areas which can be ruled out from the low carrier density. The linear magnetoresistance is also present in samples measured in a Van der Pauw geometry (not shown here), for which classical inhomogeneity models \cite{Meera2003,Meera2005} do not provide a fit \cite{Meera}. At 45 K, Abrikosov's extreme quantum limit \cite{Abrikosov1969} is certainly not fulfilled (quantum oscillations only appear at lower temperatures) and our conduction changes are orders of magnitude larger than quantum corrections to conductivity \cite{Moodera} can explain. Multiband conductivity could provide a considerable magnetoresistance, especially for similar carrier densities but very dissimilar mobilities. However, band structure calculations do not suggest very different contributions to the conduction band of Bi$_2$Te$_3$ \cite{bands}. And, most importantly, it is not possible to fit both the measured longitudinal and transverse conductivities with the same set of parameters. This observation was made before in the context of similar topological materials \cite{Andoreview}.

The Hamiltonian for our helical magnetoresistance model is given by
\begin{align}
H = \frac{\hbar^2 k^2}{2m_0} + \frac{\alpha}{\hbar} (\bm{\sigma} \times \mathbf{p}) \cdot \mathbf{e_z} + \bm{m} \cdot \bm{\sigma} + \lambda k^3 \cos \left( 3\theta \right) \sigma_z.
\label{eq:hamiltonian}
\end{align}
This Hamiltonian consists of a quadratic mass term, the Rashba spin-orbit interaction with the Pauli matrices $\bm{\sigma}=\left(\sigma_x,\sigma_y,\sigma_z \right)$ and the unit vector in the perpendicular direction $\mathbf{e_z}$, a Zeeman term with $\bm{m}=g \mu_B \bm{B}$, where $g$ is the spin Land\'e factor, $\mu_B$ the Bohr magneton and $\bm{B}$ the applied magnetic field, and a hexagonal warping term \cite{Fu2009,Adroguer} with warping parameter $\lambda$ and angle $\tan\theta=\frac{k_y}{k_x}$. The warping term applies to some materials of interest, in particular to Bi$_2$Te$_3$ used for our Hall device. The momentum is given by $k=\left| \bm{k} \right|=\left| \bm{p} \right|/\hbar$. We note that the Zeeman term can also be due to an exchange field rather than an external magnetic field. The spin-orbit term can be replaced by the Dresselhaus spin-orbit interaction, but, apart from a spin rotation of 90\textdegree{} in-plane, this does not change our results. Due to the spin-orbit interaction, the energy bands are spin-split. The separate bands are denoted by the $\pm$ indices in the following. In the absence of the quadratic mass term, the $\pm$ indices would indicate the upper and lower half of the Dirac cone (e.g. for topological insulators, where $\alpha=\hbar v_F$).

We now focus on a magnetic field applied in the out-of-plane direction, $\bm{B}=B_z \mathbf{e_z}$. The energy values of the system are $E_\pm = \beta k^2 \pm \sqrt{\alpha^2 k^2 + m_{\theta}^2}$, where $\beta = \frac{\hbar^2}{2m_0}$ and $m_{\theta}$ can be seen as the combined effect of the magnetic field and warping, $m_{\theta} = m_z + \lambda k^3 \cos 3\theta$. The corresponding spinor is $\psi_\pm = \left(  i e^{-i\theta}  \cos \phi_\pm, \pm \sin \phi_\pm \right)^T$ with $\tan \phi_\pm = {\sqrt{E_\pm - \beta k^2 - m_{\theta}}}/{\sqrt{E_\pm-\beta k^2 + m_{\theta}}}$. Within a simple Boltzmann picture, the transport lifetime $\tau_{\textrm{tr}}$ is given by $\tau_{\textrm{tr}}^{-1} \propto \int {S \left(1- \cos \theta \right) d \theta}$, where the scattering factor $S$ is determined using Fermi's golden rule, $S_\pm = |\Braket{\psi_\pm ^{\prime} | \psi_\pm}|^2$, for scattering from $\left| \psi \right>$ at 0 angle to $\left| \psi ^{\prime} \right>$ at angle $\theta$. We find $S_\pm = \left[\frac{1}{2}\alpha^2 k^2 \left(1+ \cos \theta \right)+ m_{\theta}^2 \right] / \left( \alpha^2 k^2 + m_{\theta}^2 \right)$. Note, that this expression reduces to $\frac{1}{2}(1+\cos \theta)$ if $m_{\theta} \rightarrow 0$, i.e. without applying a magnetic field and in absence of warping effects on the band structure. Neglecting the warping effect, a compact expression for the resistance is derived by multiplying the scattering factor $S_\pm$ with the Boltzmann factor $1-\cos\theta$ and integrating the result over all angles $\theta$. The resulting expression for the magnetoresistance is
\begin{align}
\frac{R(B)}{R(0)} = \frac{1+4 x^2}{1+x^2},
\label{eq:MR}
\end{align}
where $x$ is given by $x=\frac{E_z}{E_{\textrm{SO}}}$ and can be seen as a competition between the Zeeman energy $E_z=g \mu_B B$ and the spin-orbit energy at the Fermi energy $E_{\textrm{SO}}=\alpha k_F$. The difference between the zero field limit and the large field limit is a factor of 4 indeed.

We will now apply our helical magnetoresistance model to the data. For the case when the quadratic mass term dominates over the Zeeman and spin-orbit energy, the Fermi energy is given by $E_F= \beta k_F^2$ where $k_F$ follows from the mentioned Hall carrier density. Assuming a literature value for the band mass of the conduction band of Bi$_2$Te$_3$ of $\beta=$ 45 eV\AA$^2$ \cite{Liu,Sengupta} in the direction of the planes, and assuming a much larger value for the out-of-plane effective mass, the Fermi energy is found to be $E_F=$ 75 meV. Fitting the helical resistance model to the data in Fig. 2d provides the best least square fit for a zero-field mobility $\mu=$ 1750 cm$^2$V$^{-1}$s$^{-1}$ and a Zeeman-to-spin-orbit ratio $\frac{g \mu_B}{\alpha k_F}=$ 0.10 T$^{-1}$. This means that the Zeeman energy and spin-orbit energy are equal in magnitude for a magnetic field of 10 T. Assuming a $g$-factor of 12 (which could be sample specific, judging from the range of literature values \cite{gfactor}), a spin-orbit energy parameter $\alpha=0.17$ eV\AA{} is obtained. As expected, the spin-orbit interaction is much weaker in the bulk than in the Dirac cone of Bi$_2$Te$_3$. In the supplementary information we fit the helical magnetoresistance model to measurement on a 70 nm thin film of Bi$_2$Te$_3$, grown by molecular beam epitaxy. The obtained spin-orbit energy for the film is slightly higher than for the thicker flake. Note, that the zero-field mobility of $\mu=$ 1750 cm$^2$V$^{-1}$s$^{-1}$ does not warrant Shubnikov-de Haas oscillations of these bulk carriers at the largest magnetic field that we can obtain, since the helical magnetoresistance reduces this mobility with a factor of 4 above 10 T. 

With just two fitting parameters a decent fit is obtained. The fit is slightly refined by incorporating a modest warping parameter $\lambda=20$ eV\AA$^3$, see the inset of Fig. 2c. Indeed, the spin-orbit energy and Zeeman energy are small with respect to the Fermi energy, warranting the estimate for $E_F$ and the use of a one-band fit to the data. Only at the highest fields, the carrier densities are significantly different, but both mobilities have saturated towards the same value so that one crosses over again to the one-band scenario. For smaller Fermi energies, the different carrier densities for both spin-split bands would have to be taken consistently into account. In order to illustrate the broken spin helicity for the obtained fit parameters, Fig. 3 shows the dispersion relations, Fermi surfaces, and scatter factors. It is clear that for 30 T perpendicular field, the scatter factor has become almost isotropic. The influence of the small warping parameter can only be distinguished in the out-of-plane spin component $\Braket{\psi|\sigma_z|\psi}$ of Fig. 3b. Since warping enters the Hamiltonian as an effective magnetic field, it is not surprising that the effect of warping is to offset the magnetic field axis, as illustrated in Fig. 2c. Warping thereby renders the magnetoresistance already linear at low magnetic fields.

\begin{figure}
\includegraphics[width=\textwidth]{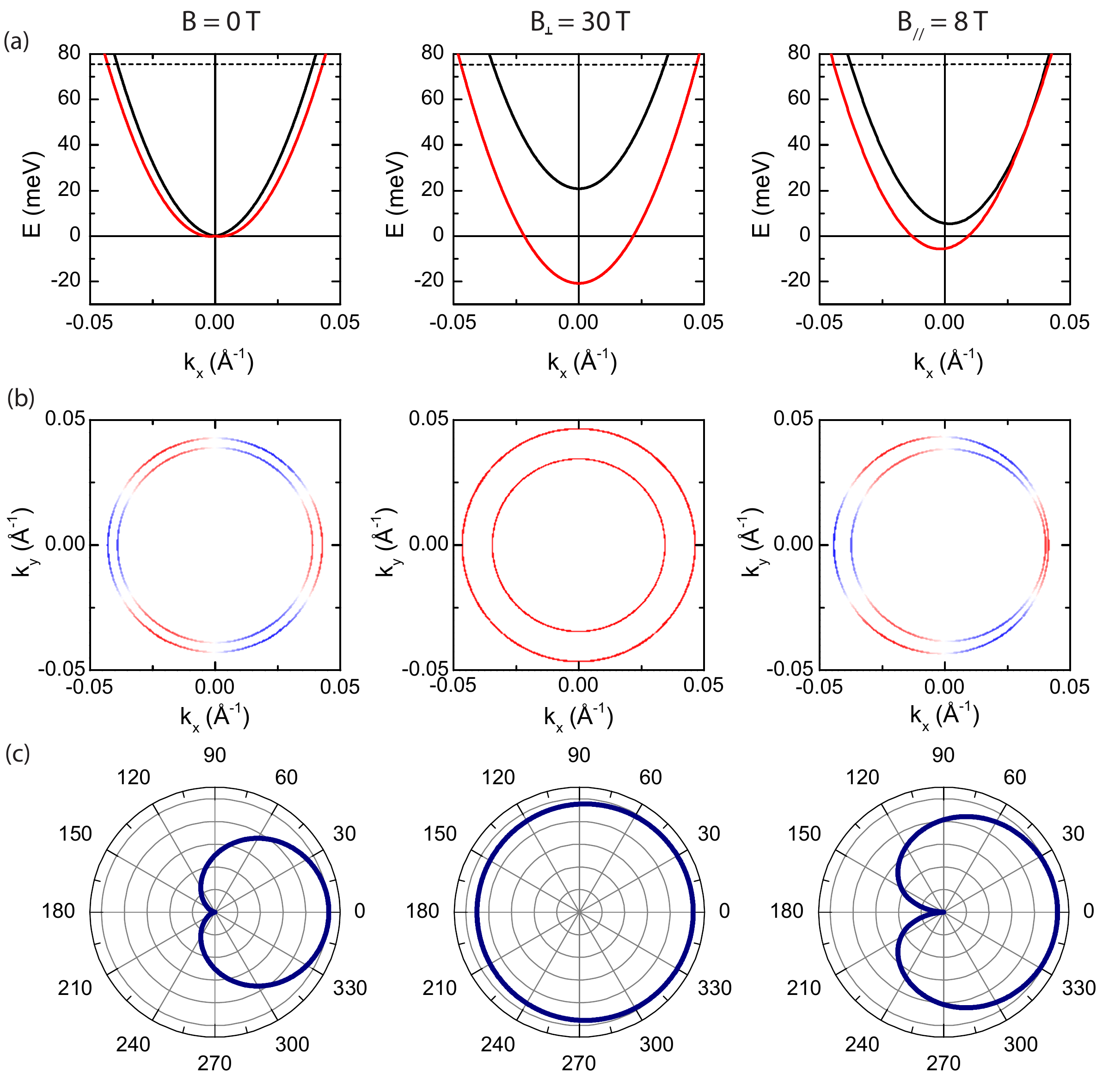}
\caption{\textbf{| Band structure and spin helicity.} (a) Dispersion relation belonging to Eq. (1) with the parameters obtained from the fit in Fig. 2: $\alpha=0.17$ eV\AA{}, $\beta=$ 45 eV\AA$^2$, $\lambda=20$ eV\AA{}$^3$, $g=12$. The $E_{\pm}$ are plotted in black and red respectively and the Fermi energy at 75 meV is indicated with a dashed line. From left to right the magnetic field is respectively 0, 30 T perpendicular to the surface, and 8 T parallel to the surface (but perpendicular to the current). (b) Fermi sheets belonging to the same set of parameters. The color (blue for negative and red for postive) indicates the out-of-plane component of the spin. The small (maximally about 23\%) hexagonal out-of-plane spin structure at $B=0$ T and $B_{//}=8$ T originates from the warping term. At 30 T perpendicular field, the spin is pointing almost completely out of plane. (c) The corresponding scatter factors, $S$, which are $1+\cos \theta$ for $B=0$ and close to isotropically 1 for $B_{\perp}=30$ T (all scatter plots use the same scale, being 1 for $\theta=0$).}  
\label{fig:3}
\end{figure}

Also for an in-plane magnetic field, the spinor eigenfunctions explicitly depend on field. An in-plane field introduces extra terms in the vector potential, but these can be gauged away. However, the Zeeman and warping terms in the spinor eigenfunctions cannot. In the supplementary information we derive the expressions for magnetoresistance for an in-plane field. Depending on the anisotropy of the $g$-factor of a material, the in-plane magnetoresistance is expected to be of similar order of magnitude. This is also visible from our measurements on bulk Bi$_2$Te$_3$, Fig. 2c, where the in-plane magnetoresistance (the field is in-plane, but perpendicular to the applied current) is only about a factor of 2 smaller than the magnetoresistance for perpendicular field.  

The magnetoresistance from broken helicity is large for Zeeman energies or exchange fields of the order of the spin-orbit energy. In topological insulators, where $\alpha=\hbar v_F$, this spin-orbit energy is very large and magnetoresistance will only appear at very low Fermi energies, close to the Dirac point, for practical values of the magnetic field. But for many two-dimensional electron gases or even bulk Rashba systems with weak to moderate spin-orbit interaction, helical magnetoresistance is relevant. The large and linear magnetoresistance in topological insulator materials \cite{He,Qu,Tang,Wang,WWang} should therefore not be taken as evidence for topological surface state transport, but rather points towards a spin-orbit splitting in the bulk or at the trivial surface states in these materials. The simple expression, Eq. (2), for the helical magnetoresistance can be used to analyze a wide class of systems. Examples of current interest are the LaAlO$_3$-SrTiO$_3$ interface with Rashba spin-orbit coupling \cite{Caviglia,Dil} and bulk Rashba systems such as BiTeI \cite{Tokura}. 

We note that several factors could reduce the magnitude of the helical resistance. The most obvious factor is interband scattering between the spin-split bands. When interband scattering dominates over intraband scattering, the magnetoresistance would vanish. For the conduction band of Bi$_2$Te$_3$ angle resolved photoemission has revealed that interband scattering is relatively weak in the comparable Bi$_2$Se$_3$ system \cite{Valla}. Throughout our simple Boltzmann model we have assumed that the spin will relax into the eigenstate of the final state wavefunction. This assumption generally holds when the elastic scatter time is much longer than the spin precession time. For very diffusive systems a D'yakanov-Perel type spin scattering \cite{DP} would therefore reduce the magnetoresistance. Factors that would enhance the magnetoresistance above a factor of 4 can also naturally occur. Especially in warped geometries scattering occurs preferentially along certain directions. The more backward directed the scattering is, the stronger the topological protection factor. Also multiband effects, e.g. relevant for field induced band depletion at pinned Fermi surfaces, can further enhance the helical magnetoresistance.

During the submission process of our manuscript a theoretical work was posted online \cite{Ozturk2014} which is consistent with the model presented here, but it is restricted to topological surface states, for which we showed the effect to be small. We acknowledge useful discussion with Meera Parish, Marieke Snelder, and Martin Stehno. This work is supported by the Netherlands Organization for Scientific Research (NWO), by the Dutch Foundation for Fundamental Research on
Matter (FOM) and by the European Research Council (ERC).

\end{document}